\documentclass[aps,pre,twocolumn,showpacs,superscriptaddress,groupedaddress]{revtex4}  
\pdfoutput=1

\usepackage{graphicx}  
\usepackage{dcolumn}   
\usepackage{bm}        
\usepackage{amssymb}   
\usepackage{amsmath}
\usepackage{color}

\begin{document}

\widetext

\title{Bass-SIR model for diffusion of new products}

\author{Gadi Fibich}%
\email{fibich@tau.ac.il} \affiliation{Department of Applied Mathematics, Tel Aviv University, Tel Aviv 69978, Israel}

\date{\today}

\begin{abstract}
We consider the diffusion of new products in social networks, where consumers who adopt the product can later ``recover'' and stop influencing others to adopt the product.
We show that the diffusion is not described by the SIR model, but rather by a novel model, the 
Bass-SIR model, which combines the Bass model for diffusion of new products with the SIR model for epidemics.
The phase transition of consumers from non-adopters to adopters is described by a non-standard Kolmogorov-Johnson-Mehl-Avrami model, in which  clusters growth is limited by adopters' recovery. 
Therefore, diffusion in the Bass-SIR model only depends on the local structure of the social network,
 but not on the average distance between consumers. Consequently, 
unlike the SIR model, a small-worlds structure has a 
negligible effect on the diffusion. 
Surprisingly, diffusion on scale-free networks is nearly identical to that on Cartesian ones.  
\end{abstract}

\maketitle

 Diffusion through social networks concerns the spreading of ``items'' ranging from diseases and computer viruses to rumors, information, opinions, technologies and innovations, and has attracted the attention of researchers  
in physics, mathematics, biology, computer science, social sciences, economics, and management science~\cite{Rogers-03,Jackson-08,Strang-98,Albert-00,Pastor-Satorras-01,Anderson-92}. 
In marketing, diffusion of new products is a fundamental problem~\cite{Mahajan-93}.
Ideally, firms would like to be able to predict future sales of a new product, its market potential, and the impact of various promotional strategies, 
  based on sales data from the first few months.

 The first mathematical model of diffusion of new products
was proposed in~1969 by Bass~\cite{Bass-69}. 
The Bass model inspired a
huge body of theoretical and empirical research
(in 2004 it was named one of the 10~most-cited papers in the 50-year
history of Management Science~\cite{Hopp-04}),
in diverse areas
such as retail service, industrial technology, agriculture,
and in educational, pharmaceutical, and consumer-durables
markets~\cite{Mahajan-93}.
In all these studies, however, it was assumed that once a consumer adopts the product, he influences other non-adopters to adopt (or disadopt) the product at all later times. 
More often than not, however, adopters ``{\em recover}'' from influencing other people  after some time.
For example, it was recently observed that new installations of solar photovoltaic (PV) systems are strongly 
influenced by the presence of nearby previously installed systems, but 
the effect of nearby PV systems decays after several months~\cite{Graziano-15}.  

In this Letter, we study the diffusion of new products when adopters are allowed to recover. 
This problem cannot be analyzed using the SIR model~\cite{SIR}, since in this model all the
{\em external adopters}, i.e., those who were not influenced by previous adopters  (``patients zero''), exist at $t=0$, 
which is not the case for new products.
 Therefore, we introduce a novel model, the Bass-SIR model, 
which allows for an on-going creation of external adopters. 
We show that this difference in the generation of external adopters is not a technical issue,
as it leads to a completely different diffusion dynamics.

To understand the effect of the network structure in the Bass-SIR model, we introduce 
 a nonstandard Kolmogorov-Johnson-Mehl-Avrami (KJMA) model for phase transitions, in which 
 clusters growth is limited by adopters' recovery.
The KJMA model with recovery may also be relevant to other problems in physics, 
such as algorithmic self-assembly of DNA tiles~\cite{Winfree-98}, where  ``recovery'' corresponds to an assembly error.

{\em Discrete Bass-SIR model.---} Consider a new product which is introduced at time $t=0$ to a market  with $M$~potential consumers.  The consumers   
belong to a social network  which is represented by an undirected graph,
such that if consumers~$i$ and~$j$ are connected, they can influence each other to adopt the product. As in the Bass model~\cite{Bass-69}, 
 if consumer~$j$ did not adopt the product by time~$t$, 
his probability to adopt (and thus become a contagious adopter) is~\cite{OR-10} 
\begin{subequations}
   \label{eq:ABM-Bass-SIR}
\begin{equation}
\label{LinearAdoptionRates-with-recovery}
\text{Prob}{j~\text{adopts in}\choose {(t,t+\Delta t)}} = \left(p+q \frac{i_j(t)}{k_j}\right) \Delta t+o(\Delta t) 
\end{equation}
as $\Delta t\to0$,
 where~$i_j(t)$ is the number of contagious adopters connected to~$j$ at time~$t$, 
and $k_j$ is the number of social connections (``degree'') of~$j$.
 The parameters~$p$ and~$q$ describe the likelihood of a consumer to
adopt the product due to {\em external influences} by mass
media or commercials, and due to {\em internal influences} by
contagious adopters to which he is connected (``word of mouth''),
respectively. In physical contexts, such influences correspond to an external source term and a drift term, respectively. In epidemics, such influences correspond to animal to human and 
human to human infections, respectively.
 The magnitude of internal influences increases linearly with the number~$i_j$  
 of contagious adopters connected to~$j$, and is normalized by~$k_j$ so that
 regardless of the network structure, the maximal internal influence that~$j$ can experience (when all his social connections are contagious adopters) is~$q$.

Unlike  previous Bass models, we do not assume that adopters remain contagious forever. Rather, as in the SIR model~\cite{SIR}, we assume that 
the probability of an adopter who was  contagious  at time~$t$ to become 
non-contagious (``recover'') in~$(t,t+\Delta t)$ is
\begin{equation}
\label{eq:recover-rate}
\text{Prob}{j~\text{recovers in} \choose {(t,t+\Delta t)}} = r \Delta t+o(\Delta t)
\end{equation}
as $\Delta t\to0$, where~$r$ is the recovery parameter. 
\end{subequations}
Since~\eqref{LinearAdoptionRates-with-recovery} and~\eqref{eq:recover-rate} 
come from the discrete Bass and SIR models, respectively, 
we refer to~\eqref{eq:ABM-Bass-SIR} as the {\em discrete Bass-SIR model}.   

We denote by~$S(t)$, $I(t)$, and $R(t)$ the fraction of non-adopters (``{\em susceptible}''), contagious adopters (``{\em infected}''), and
non-contagious adopters (``{\em recovered}'') at time~$t$, respectively.
The fraction of adopters (contagious and recovered) is denoted by
 $f = I+R = 1-S$.
 Since the product is new, initially all consumers are non-adopters, and so 
$S(0) = 1$ and $f(0) = I(0) = R(0) = 0$.

{\em Non-spatial (complete) networks.---} When all $M$~consumers are connected to each other, then $i_j(t) = M \cdot I(t)$ is the number of contagious adopters 
in the market
and  $k_j = M-1$. 
As $M \to \infty$,  
the aggregate (macroscopic) diffusion dynamics is governed by 
\footnote{This result was proved in~\cite{Niu-02} for $r=0$.  In addition, the recovery rate~\eqref{eq:recover-rate} corresponds to 
$I' = -r I$.}
\begin{subequations}
\label{eq:Bass_SIR}
\begin{equation}
\label{eq:Bass_SIR_eq}
 S'(t) = -S (p + q I), \quad I'(t) = S (p + q I)-r I, \quad  R'(t) = r I,
\end{equation}
\begin{equation}
\label{eq:Bass_SIR_ic}
S(0) = 1, \quad I(0) = 0, \quad R(0)  = 0,
\end{equation}
\end{subequations}
where $' = \frac{d}{dt}$.
In the absence of recoveries ($r=0$), $R=0$ and $f=I = 1-S$, and so eqs.~\eqref{eq:Bass_SIR} reduce 
to the original Bass model~\cite{Bass-69}
$$
    f'(t) =  (1-f)(p + q f), \qquad f(0)=0.
$$
Solving this equation yields the well-known Bass formula 
\begin{equation}
\label{eq:f_Bass}
f_{\rm Bass}(t) = \frac{1-e^{-(p+q)t}}{1+(q/p)e^{-(p+q)t}}.
\end{equation}
Similarly, when $p=0$, 
eqs.~\eqref{eq:Bass_SIR_eq} reduce
to the SIR model~\cite{SIR} 
$$
 S'(t) = -qS I, \quad I'(t) = q S  I-r I , \quad  R'(t) = r I ,
$$
which is typically supplemented by the initial conditions 
$$
S(0) = 1-I_0, \quad I(0) = I_0>0, \quad R(0)  = 0,
$$
where $I_0$ is the fraction of contagious adopters at $t=0$. 
 Therefore, we refer to~\eqref{eq:Bass_SIR} as 
the {\em continuous Bass-SIR model}.

\begin{figure}[ht!]
\begin{center}
\scalebox{0.5}{\includegraphics[trim= 2cm 14cm 0cm 8cm,clip]{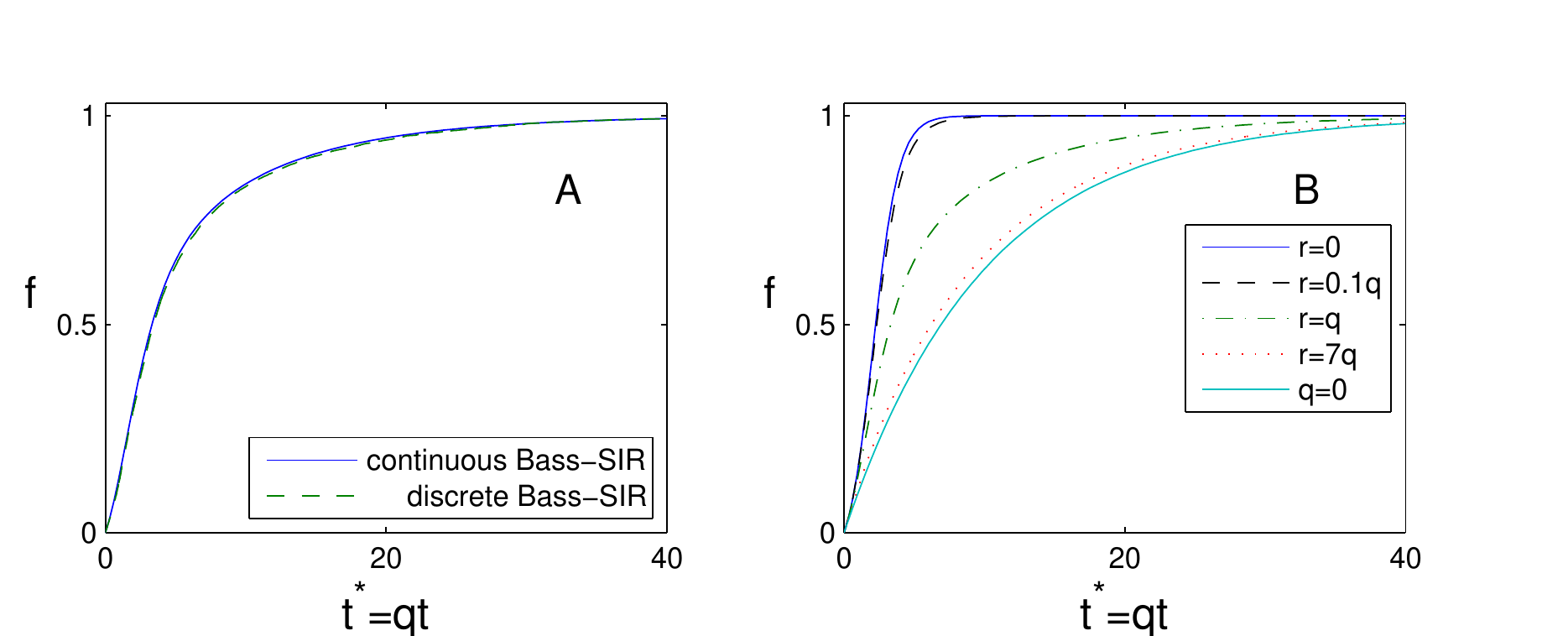}}
\caption{Fraction of adopters in the Bass-SIR model on a nonspatial network, as a function of $t^* = qt$. Here $p=0.01$ and  $q= 0.1$. 
A)~Agreement between the continuous model~\eqref{eq:Bass_SIR} [solid] and a single simulation of the discrete model~\eqref{eq:ABM-Bass-SIR} with~$M=10,000$ [dashes]. Here $r=0.1$.  
B)~The continuous model with $r=0, 0.1q, q$, and~$7q$. Here $r=0$ is $f_{\rm Bass}$, see~\eqref{eq:f_Bass},
 and $q=0$ is~$f = 1-e^{-p t}$. 
}
\label{fig:cellular_nonspatial_main}
\end{center}
\end{figure}

   Fig.~\ref{fig:cellular_nonspatial_main}A demonstrates the 
agreement between the discrete and continuous Bass-SIR models
on a nonspatial network. 
Fig.~\ref{fig:cellular_nonspatial_main}B shows 
the dependence of~$f(t)$, the fraction of adopters, on~$r$. 
 When $r \ll q$, adopters have sufficient time to influence their social contacts before they become non-contagious. Hence,
  the effect of recovery is small, and diffusion is only slightly slower than in the absence of recovery,
i.e., $f(t;p,q,r) \approx f(t;p,q,r=0)=  f_{\rm Bass}$, see~\eqref{eq:f_Bass}.
As~$r$ increases, internal influences persist for shorter times, hence diffusion becomes slower. 
  Therefore, $f$~is monotonically decreasing in~$r$.
In particular,   
  when $r \gg q$, adopters have little time to influence their social contacts before they become non-contagious. Therefore, internal influences effectively disappear, and diffusion is driven by purely-external adoptions,
i.e., $f(t;p,q,r) \approx f(t;p,q=0) = 1-e^{-pt}$. 
In particular, unless $r \ll q$, 
neglecting recovery (i.e., using the Bass model and not the Bass-SIR model) leads to 
inaccurate results.

.

{\em Cartesian networks.---} To analyze the effect of a network with a spatial structure on the diffusion, we first consider periodic $D$-dimensional Cartesian networks, where each node (consumer) is connected to 
 its $2D$ nearest neighbors.  
In that case, relation~\eqref{LinearAdoptionRates-with-recovery} reads 
\begin{equation}
\label{LinearAdoptionRates-recovery}
\text{Prob}{j~\text{adopts in}\choose {(t,t+\Delta t)}} = \left(p+q \frac{i_j(t)}{2D}\right) \Delta t+o(\Delta t). 
\end{equation}
Thus, when $D=1$ the network is a circle and each consumer can be influenced by his left and right neighbors, 
when $D=2$ the network is a torus and each consumer can be influenced by his up, down, left, and right neighbors, etc. 


%
%
%
 
\begin{figure}[ht!]
\begin{center}
\scalebox{0.5}{\includegraphics[trim= 2cm 14cm 0cm 8cm,clip]{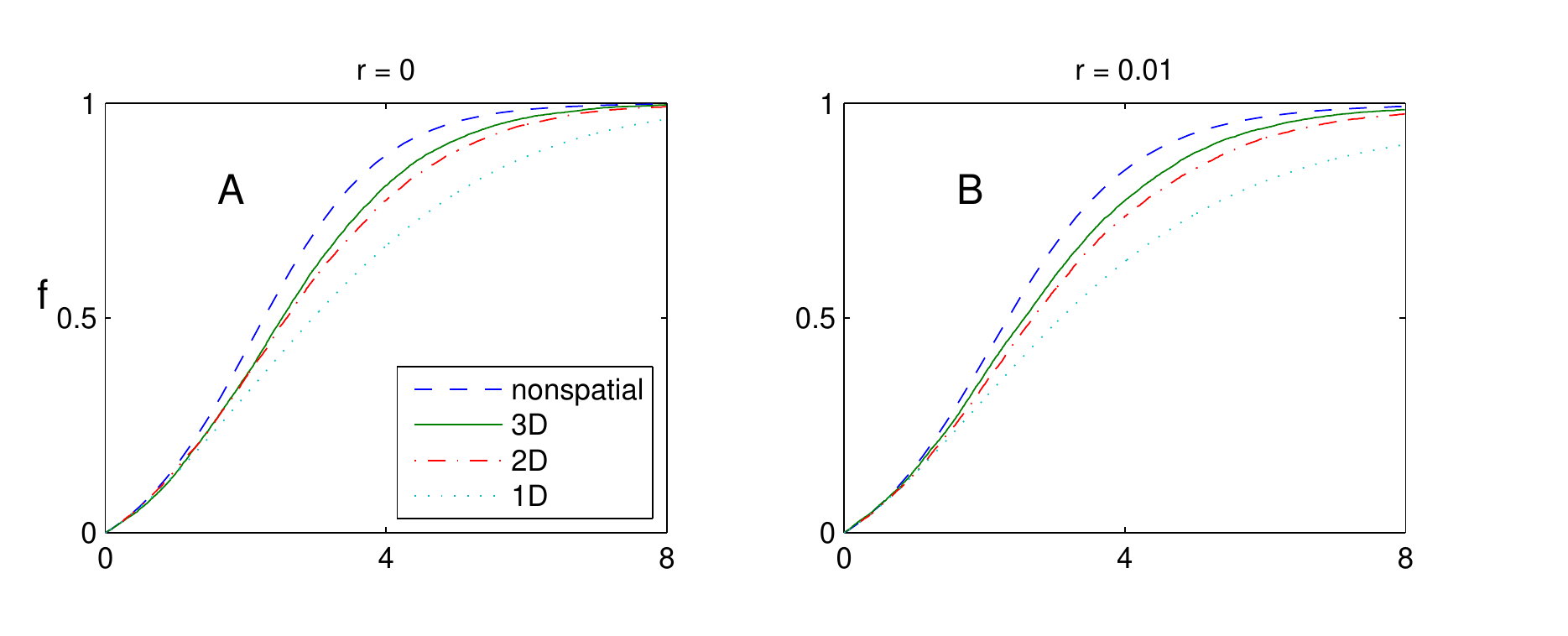}}
\caption{Fraction of adopters in the Bass-SIR model on 1D (dots), 2D (dash-dot), 3D (solid), and nonspatial (dashes) networks. 
Here $p=0.01$, $q= 0.1$, and~$M=10,000$. A)~$r=0$. B)~$r=0.1q$. C)~$r=q$. D)~$r=7q$.}
\label{fig:cellular_nonspatial_1D_2D_3D_main}
\end{center}
\end{figure}

Our simulations reveal that for given values of~$p$, $q$, and~$r$, 
diffusion in a 2D~network is faster than 
in a 1D~network but slower than in a 3D~network, which, in turn, is slower than in a nonspatial
network (Fig.~\ref{fig:cellular_nonspatial_1D_2D_3D_main}).
Note that this result is not obvious, since as a network gets more connected, the 
effect of each connection decreases, so that the maximal internal influence remains~$q$, 
see~\eqref{LinearAdoptionRates-with-recovery}. 
The differences among the four networks decrease with~$r$. 
This  is  because a larger~$r$ means  shorter internal effects, hence a 
weaker dependence on the network structure.

A priori, it may seem that diffusion becomes faster with~$D$, 
because for a Cartesian network with $M$~consumers, the average distance between consumers decreases as~$M^{1/D}$. 
If diffusion depends on the average distance between consumers, however, then increasing the population size should slow down the fractional adoption, so that 
$\lim_{M\to \infty} f(t) = 0$. This, however, is not the case, since 
$\lim_{M\to \infty} f \ge 1-e^{-pt}$. 

\begin{figure}[ht!]
\begin{center}
\scalebox{0.5}{\includegraphics[trim= 2cm 7cm 0cm 8cm,clip]{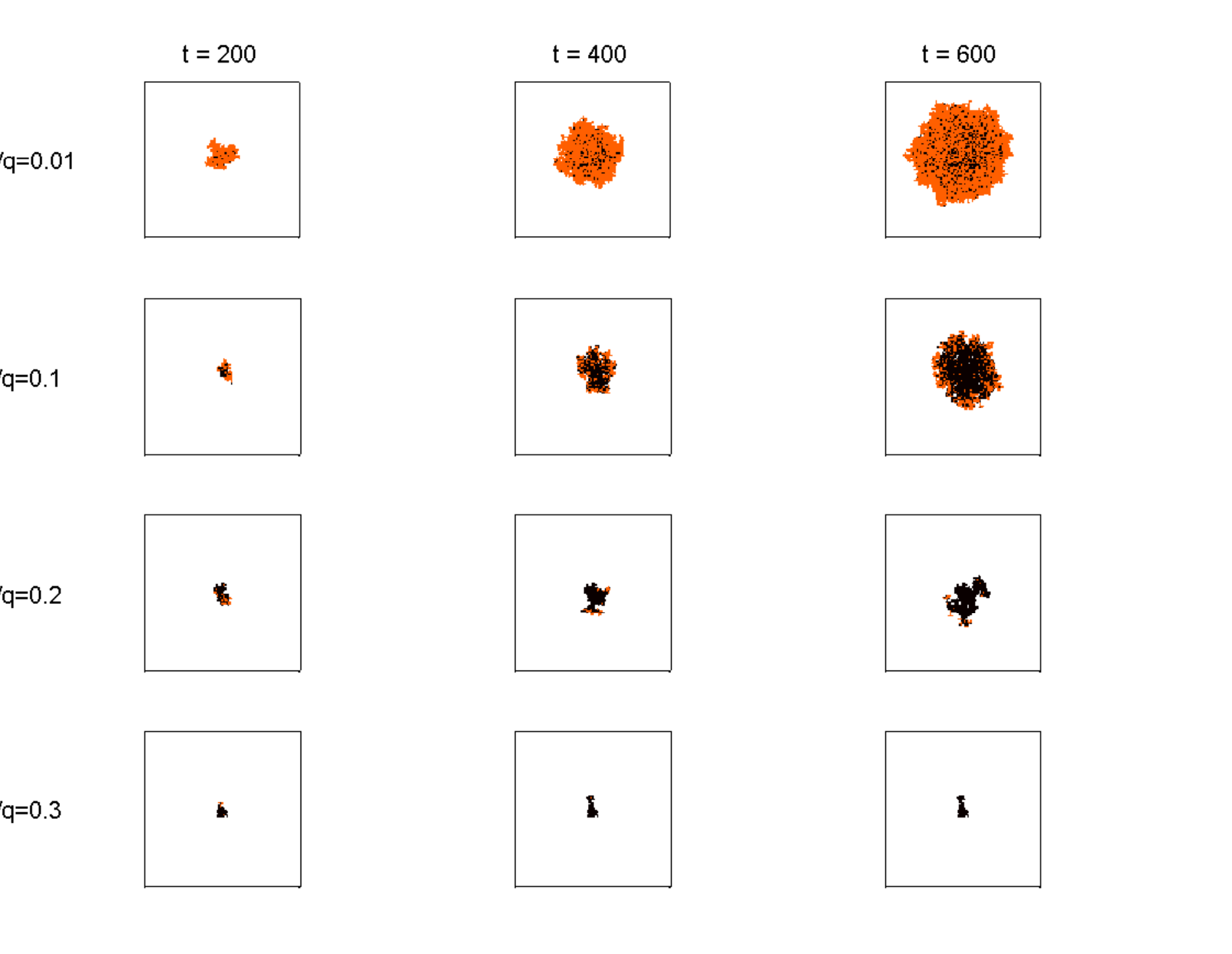}}
\caption{Typical evolution of a single cluster on a 2D network. Here $p=0$, $q=0.1$, and there is a single contagious adopter at $t=0$. Each row corresponds to a different value of~$r$. Contagious and recovered adopters are marked by orange and black pixels, respectively.  }
\label{fig:cluster_2D_dynamics_main}
\end{center}
\end{figure}

{\em Kolmogorov-Johnson-Mehl-Avrami (KJMA) model.---} To understand the effect of the network structure,  it is useful to visualize
the diffusion process as an on-going random creation of external adopters (``seeds''). Once created, each seed expands 
through internal adoptions into a cluster of adopters, and expanding clusters can merge into  larger clusters.    
This is nothing but the KJMA model for phase transitions~\cite{Kolmogorov-37,Johnson-39,Avrami-39}
from non-adopters to adopters. 
Unlike its standard applications in physics (but as in algorithmic self-assembly of DNA tiles~\footnote{Here, recovery corresponds to an assembly error~\cite{Barish-09}.}), 
in the Bass-SIR model 
the evolution of clusters can be more complex, because of the recoveries.
To see that, in Fig.~\ref{fig:cluster_2D_dynamics_main} we simulate the evolution of single cluster in a 2D network, by
placing a single contagious adopter at $t=0$, and setting $p=0$ in~\eqref{LinearAdoptionRates-with-recovery} so that
all subsequent adoptions are purely-internal. When~$r$ is sufficiently small, 
clusters expand as squares/circles, whose radius increases with time
(top row). As~$r$ increases, the cluster expands more slowly, 
the fraction of recovered adopters (out of all adopters) increases, and contagious adopters are mostly concentrated near the cluster surface (second raw). As~$r$ further increases (third row), 
some adopters on the cluster boundary recover before they lead to new adoptions.
As a result, clusters evolve into irregular shapes. 
As~$r$ further increases (fourth row), the cluster ceases to expand after some time, once all of its adopters became non-contagious. 
%

%
%

Since external adoptions are independent of the network structure, the KJMA model 
implies that {\em the network structure affects the diffusion by affecting the average rate at which clusters expand}. 
We can use this insight to explain the results of Fig.~\ref{fig:cellular_nonspatial_1D_2D_3D_main}, as follows.
The average radius~$\rho$ of clusters of size~$N$ scales as~$N^{\frac1D}$, hence their average surface area 
(i.e., the number of adopters on the cluster surface which are in direct 
contact with non-adopters) scales as
$\rho^{D-1} \sim N^{\frac{D-1}{D}}$. 
Therefore, the cluster expansion rate scales as $q N^{\frac{D-1}{D}}/D$, see~\eqref{LinearAdoptionRates-recovery}.  
Hence, the higher $D$~is,
the faster the cluster expansion is~\cite{OR-10}.

{\em Small-worlds network.---} The structure of real-life social networks is different from that of Cartesian networks. Watts and Strogatz~\cite{Watts-98} suggested that social networks have a small-worlds structure, whereby most connections are local, 
but there are also some random long-range connections. 
They showed that even a small fraction of long-range connections can lead to a dramatic reduction in the average distance between nodes. As a result,  
epidemics spread much faster on networks with a small-worlds structure.

The acceleration of diffusion by a small-worlds structure should be maximal in the 1D~model,
because for a given~$M$, the average distance is maximal in 
the 1D~model. To induce a ``{\em 5\% small-worlds structure}'' in the 1D~model,
we add a link between any two nodes with probability~$0.05/M$, so that the average graph degree increases
from~2 to~2.05.
If, as a result, $j$~is connected to $k_j>2$ other consumers, we change the internal effect of each of these consumers on~$j$
from~$q/2$ to~$q/k_j$, in accordance with~\eqref{LinearAdoptionRates-with-recovery}.

\begin{figure}[ht!]
\begin{center}
\scalebox{0.5}{\includegraphics[trim= 2cm 7cm 0cm 8cm,clip]{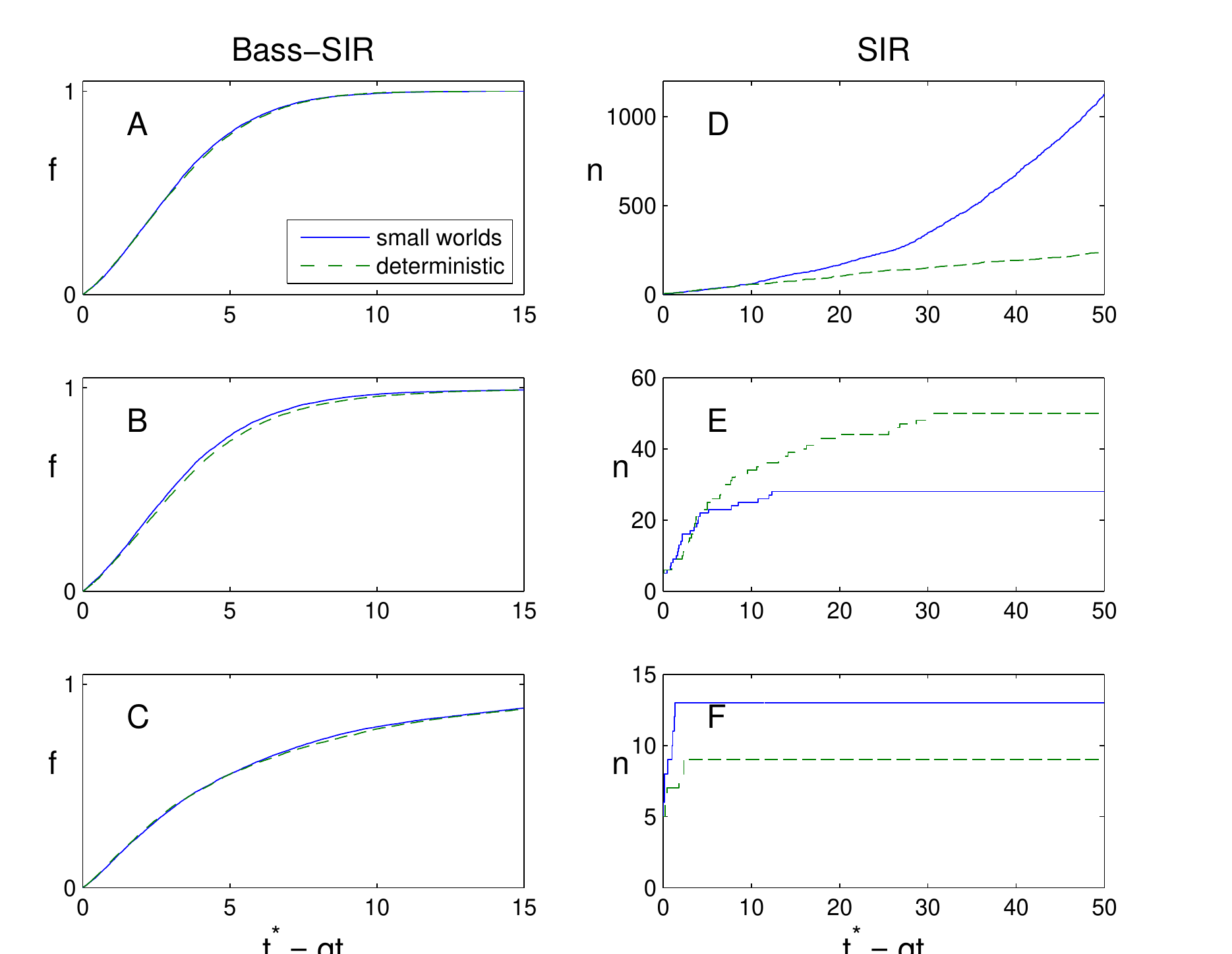}}
\caption{Diffusion on a 1D~network with (solid) and without (dashes) a small-worlds structure. Here $ q= 0.1$ and $M=10,000$. (A)--(C)~Fraction of adopters in the discrete Bass-SIR model. Here $p=0.01$, and there are no adopters at $t=0$.  A)~$r=0$. B)~$r=0.1q$. C)~$r=q$. (D)--(F)  Number of adopters in the discrete SIR model.
Here $p=0$, and diffusion starts from 5 randomly-chosen contagious adopters at $t=0$.
 D)~$r=0$. E)~$r=0.1q$. F)~$r=q$. }
\label{fig:cellular_small_worlds_main}
\end{center}
\end{figure}

Our simulations reveal that
{\em the addition of a small-worlds structure has a negligible effect on diffusion of new products} 
(Fig.~\ref{fig:cellular_small_worlds_main}A--C). 
This is because a small-worlds structure reduces the average distance between agents. This global network property has a negligible effect, however,  
on diffusion of new products, 
which depends on the growth rate of a cluster, 
which, in turn, depends on local properties of the network (such as the grid dimension~$D$). Indeed, roughly speaking, 
if in the absence of a small-worlds structure  a certain cluster reaches at time~$t$ a size of~$N(t)=20$, 
then adding a ``5\% small-worlds structure'' would 
increase $N(t)$ at most to~$21$.

To show that our results are not inconsistent with~\cite{Watts-98},
in Fig.~\ref{fig:cellular_small_worlds_main}D--F  we repeat these simulations for the SIR model 
on the same 1D network with 5\% small-worlds structure,
 and with the same values of~$q$ and~$r$. Thus, the only differences
from Fig.~\ref{fig:cellular_small_worlds_main}A--C 
is that we now set $p=0$ in~\eqref{LinearAdoptionRates-with-recovery}, and we let 5 randomly-chosen agents be contagious adopters at~$t=0$. 
In this case the small-worlds structure has a major effect on diffusion, in agreement with~\cite{Watts-98}.
Interestingly, while in the absence of recovery a small-worlds structure always accelerates diffusion,
in the presence of recoveries it may also slow it down (Fig.~\ref{fig:cellular_small_worlds_main}E).

{\em Scale-free networks.---}  Another popular model for social networks is that of a 
scale-free network. We constructed scale-free networks  using the 
 Barab\'asi-Albert (BA) preferential-attachment algorithm~\cite{BA-99},
in which each new node makes $m$~new links with the existing
 network nodes, such that the probability of a new node to connect to node~$i_0$ is 
$k_{i_0}/\sum_i k_i$, where $k_i$ is the degree of node~$i$.
 In the resulting scale-free network, if node~$j$ is connected to $k_j$~nodes, 
the effect of each of these nodes 
on~$j$ is~${q}/{k_j}$, see~\eqref{LinearAdoptionRates-with-recovery}.

\begin{figure}[ht!]
\begin{center}
\scalebox{0.5}{\includegraphics[trim= -2cm 14cm 0cm 8cm,clip]{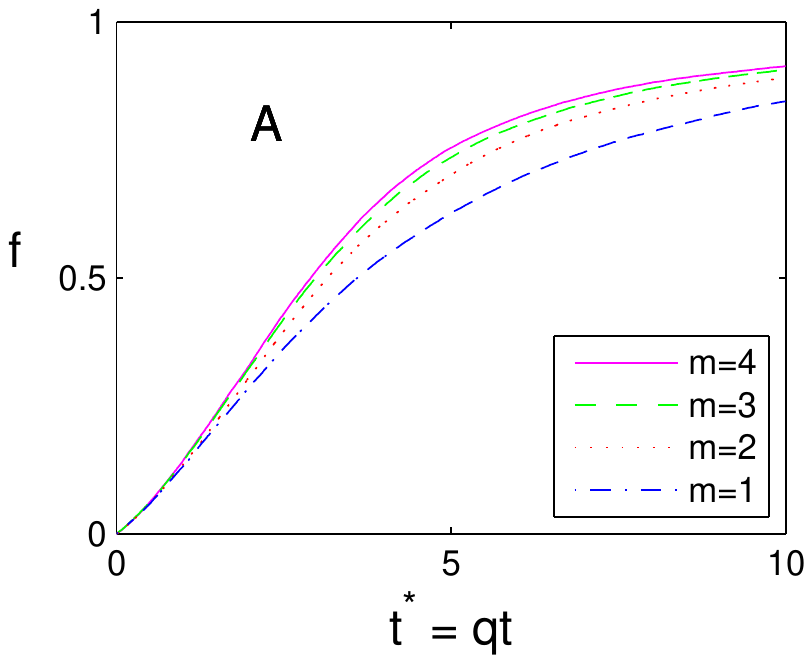}}
\scalebox{0.5}{\includegraphics[trim= 2cm 8cm 0cm 8cm,clip]{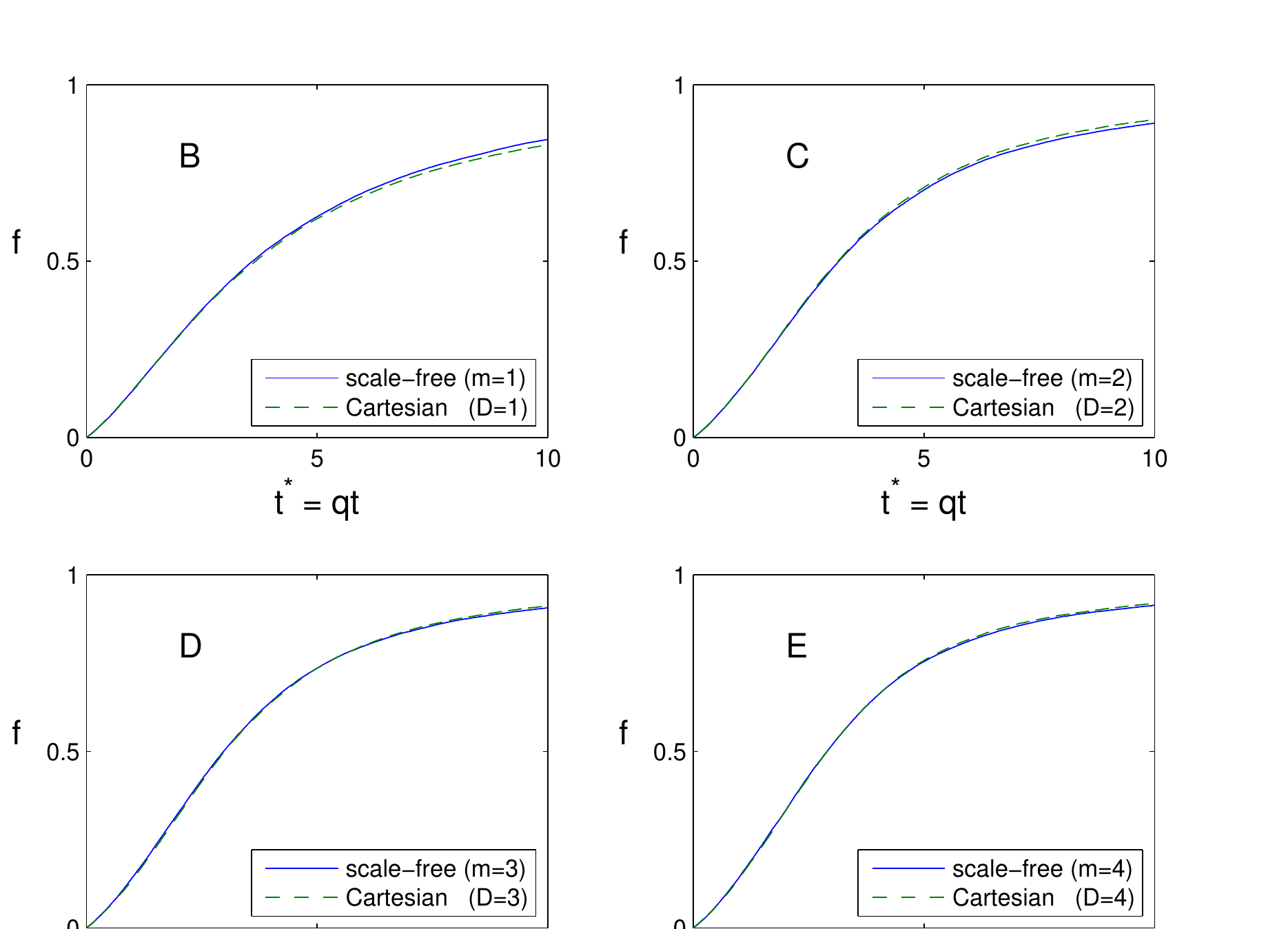}}
\caption{Fraction of adopters in the Bass-SIR model~\eqref{eq:ABM-Bass-SIR} on networks with
$p=0.01$, 
$q= 0.1$, $r= 0.5q$, and $M=50,000$. 
A)  Scale-free networks with~$m=1,2,3$, and~$4$.
(B)--(E) Scale-free (solid) and Cartesian (dashes) networks. 
B)~$m=D=1$. C)~$m=D=2$. D)~$m=D=3$. E)~$m=D=4$.
 }
\label{fig:cellular_scale_free_main_M=50000}
\end{center}
\end{figure}

 Our simulations of the discrete Bass-SIR model~\eqref{eq:ABM-Bass-SIR} on scale-free networks show that,  
as expected, the larger~$m$ is, the faster the diffusion (Fig.~\ref{fig:cellular_scale_free_main_M=50000}A). 
Surprisingly, {\em the diffusion on a scale-free network with parameter~$m$ is nearly identical to that on a Cartesian network with $D=m$} (Fig.~\ref{fig:cellular_scale_free_main_M=50000}B--E). 
This numerical observation is very surprising, since these networks are different from each other
in almost any aspect. 
Yet, for some reason, the average growth rate of clusters is nearly identical in these networks. 


%

{\em Summary.---}  Two fundamental models of diffusion in social networks are the Bass model for new products, and the SIR model for epidemics~\cite{Jackson-08}. 
To the best of our knowledge, these models have not been combined into a single model until now.

The Bass-SIR model is fundamentally different from either of these models.  
Indeed, since the Bass model does not allow for recovery, it
cannot be used for products for which recovery affects the diffusion (e.g., solar PV systems). 
The SIR model does allow for recovery. However, 
in the SIR model all external adopters exist at~$t=0$, whereas in the 
Bass-SIR model there is an on-going generation of new external adopters.  
This difference in the generation of external adopters is not a technical issue,
as it leads to 
a completely different diffusion dynamics.

The key difference between these models is that in the SIR model there is a threshold value of~$r$, above which the 
epidemics will peter out. In contrast, in the Bass-SIR model everyone eventually adopts, 
since $f(t;p,q,r) \ge f(t;p,0,r)= 1-e^{-pt}$ \footnote{ 
This does not mean that the entire population  eventually adopts the product, but rather that the  model only considers the people in the population who will ultimately adopt the product (the ``market potential'') Indeed, a key goal of the Bass model is to predict the market potential.}.

The effect of the social network structure in these two models is also very different.
Thus, in the SIR model, diffusion occurs through the expansion of a single cluster of internal 
adopters around ``{\em patient zero}''. Therefore, the key determinant of diffusion speed 
is the average distance from patient zero (or more generally, the average distance
between individuals), which is a global property of the network. 
In contrast, diffusion in the Bass-SIR model  occurs through the expansion of numerous clusters.
Therefore, the diffusion speed is determined by the growth rate of clusters, 
which depends on local properties of the network.
It is because of these differences that, e.g., 
(i)~A small-worlds structure has a large effect on diffusion in  the SIR model, but a negligible one in the Bass-SIR model. 
     (ii)~Doubling the population size roughly doubles the time~$T_{1/2}$ for 
half of the population to adopt in the SIR model, but has a negligible effect on~$T_{1/2}$ in the Bass-SIR model.  

The choice between the Bass-SIR model and the SIR model
depends on the initiation of the diffusion process. Diseases and rumors that start from a ``patent zero''  call for the SIR model. External adoptions of new products 
and external infections from mosquito bites  
are on-going processes, and thus call for the Bass-SIR model. 

Some remaining open questions concern the effect of the network structure.For example, is it true that as $M\to \infty$, the diffusion on a scale-free network becomes identical to that on a Cartesian network with $D=m$? Can we derive  macroscopic (averaged) equations for diffusion in Cartesian and scale-free networks? 
How does recovery affect the phase transition kinetics in the KJMA model?

{\em Acknowledgments.---}
We thank G. Ariel and O. Raz for useful discussions.
This research was conducted while the author was visiting the Center for Scientific Computation and Mathematical Modeling (CSCAMM) at the University of Maryland, and 
was  partially supported by the Kinetic Research Network (KI-Net) under NSF Grant No. RNMS  \#1107444.

\bibliographystyle{apsrev4-1} 
\bibliography{diffusion}

\begin{thebibliography}{20}%
\makeatletter
\providecommand \@ifxundefined [1]{%
 \@ifx{#1\undefined}
}%
\providecommand \@ifnum [1]{%
 \ifnum #1\expandafter \@firstoftwo
 \else \expandafter \@secondoftwo
 \fi
}%
\providecommand \@ifx [1]{%
 \ifx #1\expandafter \@firstoftwo
 \else \expandafter \@secondoftwo
 \fi
}%
\providecommand \natexlab [1]{#1}%
\providecommand \enquote  [1]{``#1''}%
\providecommand \bibnamefont  [1]{#1}%
\providecommand \bibfnamefont [1]{#1}%
\providecommand \citenamefont [1]{#1}%
\providecommand \href@noop [0]{\@secondoftwo}%
\providecommand \href [0]{\begingroup \@sanitize@url \@href}%
\providecommand \@href[1]{\@@startlink{#1}\@@href}%
\providecommand \@@href[1]{\endgroup#1\@@endlink}%
\providecommand \@sanitize@url [0]{\catcode `\\12\catcode `\$12\catcode
  `\&12\catcode `\#12\catcode `\^12\catcode `\_12\catcode `\%12\relax}%
\providecommand \@@startlink[1]{}%
\providecommand \@@endlink[0]{}%
\providecommand \url  [0]{\begingroup\@sanitize@url \@url }%
\providecommand \@url [1]{\endgroup\@href {#1}{\urlprefix }}%
\providecommand \urlprefix  [0]{URL }%
\providecommand \Eprint [0]{\href }%
\providecommand \doibase [0]{http://dx.doi.org/}%
\providecommand \selectlanguage [0]{\@gobble}%
\providecommand \bibinfo  [0]{\@secondoftwo}%
\providecommand \bibfield  [0]{\@secondoftwo}%
\providecommand \translation [1]{[#1]}%
\providecommand \BibitemOpen [0]{}%
\providecommand \bibitemStop [0]{}%
\providecommand \bibitemNoStop [0]{.\EOS\space}%
\providecommand \EOS [0]{\spacefactor3000\relax}%
\providecommand \BibitemShut  [1]{\csname bibitem#1\endcsname}%
\let\auto@bib@innerbib\@empty
\bibitem [{\citenamefont {Rogers}(2003)}]{Rogers-03}%
  \BibitemOpen
  \bibfield  {author} {\bibinfo {author} {\bibfnamefont {E.}~\bibnamefont
  {Rogers}},\ }\href@noop {} {\emph {\bibinfo {title} {Diffusion of
  Innovations}}},\ \bibinfo {edition} {5th}\ ed.\ (\bibinfo  {publisher} {Free
  Press},\ \bibinfo {address} {New York},\ \bibinfo {year} {2003})\BibitemShut
  {NoStop}%
\bibitem [{\citenamefont {Jackson}(2008)}]{Jackson-08}%
  \BibitemOpen
  \bibfield  {author} {\bibinfo {author} {\bibfnamefont {M.}~\bibnamefont
  {Jackson}},\ }\href@noop {} {\emph {\bibinfo {title} {Social and Economic
  Networks}}}\ (\bibinfo  {publisher} {Princeton University Press},\ \bibinfo
  {address} {Princeton and Oxford},\ \bibinfo {year} {2008})\BibitemShut
  {NoStop}%
\bibitem [{\citenamefont {Strang}\ and\ \citenamefont
  {Soule}(1998)}]{Strang-98}%
  \BibitemOpen
  \bibfield  {author} {\bibinfo {author} {\bibfnamefont {D.}~\bibnamefont
  {Strang}}\ and\ \bibinfo {author} {\bibfnamefont {S.}~\bibnamefont {Soule}},\
  }\href@noop {} {\bibfield  {journal} {\bibinfo  {journal} {Annu. Rev.
  Sociol.}\ }\textbf {\bibinfo {volume} {24}},\ \bibinfo {pages} {265}
  (\bibinfo {year} {1998})}\BibitemShut {NoStop}%
\bibitem [{\citenamefont {Albert}\ \emph {et~al.}(2000)\citenamefont {Albert},
  \citenamefont {Jeong},\ and\ \citenamefont {Barab\'asi}}]{Albert-00}%
  \BibitemOpen
  \bibfield  {author} {\bibinfo {author} {\bibfnamefont {R.}~\bibnamefont
  {Albert}}, \bibinfo {author} {\bibfnamefont {H.}~\bibnamefont {Jeong}}, \
  and\ \bibinfo {author} {\bibfnamefont {A.}~\bibnamefont {Barab\'asi}},\
  }\href@noop {} {\bibfield  {journal} {\bibinfo  {journal} {Nature}\ }\textbf
  {\bibinfo {volume} {406}},\ \bibinfo {pages} {378} (\bibinfo {year}
  {2000})}\BibitemShut {NoStop}%
\bibitem [{\citenamefont {Pastor-Satorras}\ and\ \citenamefont
  {Vespignani}(2001)}]{Pastor-Satorras-01}%
  \BibitemOpen
  \bibfield  {author} {\bibinfo {author} {\bibfnamefont {R.}~\bibnamefont
  {Pastor-Satorras}}\ and\ \bibinfo {author} {\bibfnamefont {A.}~\bibnamefont
  {Vespignani}},\ }\href@noop {} {\bibfield  {journal} {\bibinfo  {journal}
  {Phys. Rev. Lett.}\ }\textbf {\bibinfo {volume} {86}},\ \bibinfo {pages}
  {3200} (\bibinfo {year} {2001})}\BibitemShut {NoStop}%
\bibitem [{\citenamefont {Anderson}\ and\ \citenamefont
  {May}(1992)}]{Anderson-92}%
  \BibitemOpen
  \bibfield  {author} {\bibinfo {author} {\bibfnamefont {R.}~\bibnamefont
  {Anderson}}\ and\ \bibinfo {author} {\bibfnamefont {R.}~\bibnamefont {May}},\
  }\href@noop {} {\emph {\bibinfo {title} {Infectious Diseases of Humans}}}\
  (\bibinfo  {publisher} {Oxford University Press},\ \bibinfo {address}
  {Oxford},\ \bibinfo {year} {1992})\BibitemShut {NoStop}%
\bibitem [{\citenamefont {Mahajan}\ \emph {et~al.}(1993)\citenamefont
  {Mahajan}, \citenamefont {Muller},\ and\ \citenamefont {Bass}}]{Mahajan-93}%
  \BibitemOpen
  \bibfield  {author} {\bibinfo {author} {\bibfnamefont {V.}~\bibnamefont
  {Mahajan}}, \bibinfo {author} {\bibfnamefont {E.}~\bibnamefont {Muller}}, \
  and\ \bibinfo {author} {\bibfnamefont {F.}~\bibnamefont {Bass}},\ }in\
  \href@noop {} {\emph {\bibinfo {booktitle} {Handbooks in Operations Research
  and Management Science}}},\ Vol.~\bibinfo {volume} {5},\ \bibinfo {editor}
  {edited by\ \bibinfo {editor} {\bibfnamefont {J.}~\bibnamefont {Eliashberg}}\
  and\ \bibinfo {editor} {\bibfnamefont {G.}~\bibnamefont {Lilien}}}\ (\bibinfo
   {publisher} {North-Holland, Amsterdam},\ \bibinfo {year} {1993})\ pp.\
  \bibinfo {pages} {349--408}\BibitemShut {NoStop}%
\bibitem [{\citenamefont {Bass}(1969)}]{Bass-69}%
  \BibitemOpen
  \bibfield  {author} {\bibinfo {author} {\bibfnamefont {F.}~\bibnamefont
  {Bass}},\ }\href@noop {} {\bibfield  {journal} {\bibinfo  {journal}
  {Management Sci.}\ }\textbf {\bibinfo {volume} {15}},\ \bibinfo {pages} {215}
  (\bibinfo {year} {1969})}\BibitemShut {NoStop}%
\bibitem [{\citenamefont {W.J.~Hopp}(2004)}]{Hopp-04}%
  \BibitemOpen
  \bibfield  {author} {\bibinfo {author} {\bibfnamefont {e.}~\bibnamefont
  {W.J.~Hopp}},\ }\href@noop {} {\bibfield  {journal} {\bibinfo  {journal}
  {Management Sci.}\ }\textbf {\bibinfo {volume} {50}},\ \bibinfo {pages}
  {1763} (\bibinfo {year} {2004})}\BibitemShut {NoStop}%
\bibitem [{\citenamefont {Graziano}\ and\ \citenamefont
  {Gillingham}(2015)}]{Graziano-15}%
  \BibitemOpen
  \bibfield  {author} {\bibinfo {author} {\bibfnamefont {M.}~\bibnamefont
  {Graziano}}\ and\ \bibinfo {author} {\bibfnamefont {K.}~\bibnamefont
  {Gillingham}},\ }\href@noop {} {\bibfield  {journal} {\bibinfo  {journal} {J.
  Econ. Geogr.}\ }\textbf {\bibinfo {volume} {15}},\ \bibinfo {pages} {815}
  (\bibinfo {year} {2015})}\BibitemShut {NoStop}%
\bibitem [{\citenamefont {Kermack}\ and\ \citenamefont
  {McKendrick}(1927)}]{SIR}%
  \BibitemOpen
  \bibfield  {author} {\bibinfo {author} {\bibfnamefont {W.}~\bibnamefont
  {Kermack}}\ and\ \bibinfo {author} {\bibfnamefont {A.}~\bibnamefont
  {McKendrick}},\ }\href@noop {} {\bibfield  {journal} {\bibinfo  {journal}
  {Proc. Roy. Soc. Lond. A}\ }\textbf {\bibinfo {volume} {115}},\ \bibinfo
  {pages} {700} (\bibinfo {year} {1927})}\BibitemShut {NoStop}%
\bibitem [{\citenamefont {Winfree}\ \emph {et~al.}(1998)\citenamefont
  {Winfree}, \citenamefont {Liu}, \citenamefont {Wenzler},\ and\ \citenamefont
  {Seeman}}]{Winfree-98}%
  \BibitemOpen
  \bibfield  {author} {\bibinfo {author} {\bibfnamefont {E.}~\bibnamefont
  {Winfree}}, \bibinfo {author} {\bibfnamefont {F.}~\bibnamefont {Liu}},
  \bibinfo {author} {\bibfnamefont {L.}~\bibnamefont {Wenzler}}, \ and\
  \bibinfo {author} {\bibfnamefont {N.}~\bibnamefont {Seeman}},\ }\href@noop {}
  {\bibfield  {journal} {\bibinfo  {journal} {Nature}\ }\textbf {\bibinfo
  {volume} {394}},\ \bibinfo {pages} {539} (\bibinfo {year}
  {1998})}\BibitemShut {NoStop}%
\bibitem [{\citenamefont {Fibich}\ and\ \citenamefont {Gibori}(2010)}]{OR-10}%
  \BibitemOpen
  \bibfield  {author} {\bibinfo {author} {\bibfnamefont {G.}~\bibnamefont
  {Fibich}}\ and\ \bibinfo {author} {\bibfnamefont {R.}~\bibnamefont
  {Gibori}},\ }\href@noop {} {\bibfield  {journal} {\bibinfo  {journal} {Oper.
  Res.}\ }\textbf {\bibinfo {volume} {58}},\ \bibinfo {pages} {1450} (\bibinfo
  {year} {2010})}\BibitemShut {NoStop}%
\bibitem [{\citenamefont {Kolmogorov}(1937)}]{Kolmogorov-37}%
  \BibitemOpen
  \bibfield  {author} {\bibinfo {author} {\bibfnamefont {A.}~\bibnamefont
  {Kolmogorov}},\ }\href@noop {} {\bibfield  {journal} {\bibinfo  {journal}
  {Izv. Akad. Nauk SSSR Ser. Fiz.}\ }\textbf {\bibinfo {volume} {1}},\ \bibinfo
  {pages} {335} (\bibinfo {year} {1937})}\BibitemShut {NoStop}%
\bibitem [{\citenamefont {Johnson}\ and\ \citenamefont
  {Mehl}(1939)}]{Johnson-39}%
  \BibitemOpen
  \bibfield  {author} {\bibinfo {author} {\bibfnamefont {W.}~\bibnamefont
  {Johnson}}\ and\ \bibinfo {author} {\bibfnamefont {P.}~\bibnamefont {Mehl}},\
  }\href@noop {} {\bibfield  {journal} {\bibinfo  {journal} {Trans. Am. Inst.
  Min. Engin.}\ }\textbf {\bibinfo {volume} {135}},\ \bibinfo {pages} {416}
  (\bibinfo {year} {1939})}\BibitemShut {NoStop}%
\bibitem [{\citenamefont {Avrami}(1939)}]{Avrami-39}%
  \BibitemOpen
  \bibfield  {author} {\bibinfo {author} {\bibfnamefont {M.}~\bibnamefont
  {Avrami}},\ }\href@noop {} {\bibfield  {journal} {\bibinfo  {journal} {J.
  Chem. Phys.}\ }\textbf {\bibinfo {volume} {7}},\ \bibinfo {pages} {1103}
  (\bibinfo {year} {1939})}\BibitemShut {NoStop}%
\bibitem [{\citenamefont {Watts}\ and\ \citenamefont
  {Strogatz}(1998)}]{Watts-98}%
  \BibitemOpen
  \bibfield  {author} {\bibinfo {author} {\bibfnamefont {D.}~\bibnamefont
  {Watts}}\ and\ \bibinfo {author} {\bibfnamefont {S.}~\bibnamefont
  {Strogatz}},\ }\href@noop {} {\bibfield  {journal} {\bibinfo  {journal}
  {Nature}\ }\textbf {\bibinfo {volume} {393}},\ \bibinfo {pages} {440}
  (\bibinfo {year} {1998})}\BibitemShut {NoStop}%
\bibitem [{\citenamefont {Barab\'asi}\ and\ \citenamefont
  {Albert}(1999)}]{BA-99}%
  \BibitemOpen
  \bibfield  {author} {\bibinfo {author} {\bibfnamefont {A.}~\bibnamefont
  {Barab\'asi}}\ and\ \bibinfo {author} {\bibfnamefont {R.}~\bibnamefont
  {Albert}},\ }\href@noop {} {\bibfield  {journal} {\bibinfo  {journal}
  {Science}\ }\textbf {\bibinfo {volume} {286}},\ \bibinfo {pages} {509}
  (\bibinfo {year} {1999})}\BibitemShut {NoStop}%
\bibitem [{\citenamefont {Niu}(2002)}]{Niu-02}%
  \BibitemOpen
  \bibfield  {author} {\bibinfo {author} {\bibfnamefont {S.}~\bibnamefont
  {Niu}},\ }\href@noop {} {\bibfield  {journal} {\bibinfo  {journal} {Math.
  Problems Engrg.}\ }\textbf {\bibinfo {volume} {8}},\ \bibinfo {pages} {249}
  (\bibinfo {year} {2002})}\BibitemShut {NoStop}%
\bibitem [{\citenamefont {Barish}\ \emph {et~al.}(2009)\citenamefont {Barish},
  \citenamefont {Schulman}, \citenamefont {Rothemund},\ and\ \citenamefont
  {Winfree}}]{Barish-09}%
  \BibitemOpen
  \bibfield  {author} {\bibinfo {author} {\bibfnamefont {R.}~\bibnamefont
  {Barish}}, \bibinfo {author} {\bibfnamefont {R.}~\bibnamefont {Schulman}},
  \bibinfo {author} {\bibfnamefont {P.}~\bibnamefont {Rothemund}}, \ and\
  \bibinfo {author} {\bibfnamefont {E.}~\bibnamefont {Winfree}},\ }\href@noop
  {} {\bibfield  {journal} {\bibinfo  {journal} {Proc. Natl. Acad. Sci.
  U.S.A.}\ }\textbf {\bibinfo {volume} {106}},\ \bibinfo {pages} {6054}
  (\bibinfo {year} {2009})}\BibitemShut {NoStop}%
\end{thebibliography}%

\end{document}